\documentstyle[epsfig,twoside,fleqn,espcrc2]{article}

\newcommand{\AmS}{{\protect\the\textfont2
  A\kern-.1667em\lower.5ex\hbox{M}\kern-.125emS}}
\newcommand {\ignore}[1]{}

\title{                 Atmospheric neutrino data : 
Active-Active $\times$ Active-Sterile oscillations
                       }
\author{O. L. G. Peres\thanks{To Appear in Proceedings of the XTH 
Int. Symposium on Very High Energy Cosmic Ray Interactions, 
Laboratory Nazionali del  Gran Sasso, Assergi, Italy, July 12-17 1998. 
I thank FAPESP by the finantial support.
} 
\address{Instituto de F\'{\i}sica Corpuscular - C.S.I.C.\\
   Departament de F\'{\i}sica Te\`orica, Universitat de Val\`encia\\
   46100 Burjassot, Val\`encia, Spain}}

\begin{document}
\begin{abstract}
I summarize here the results of a global 
fit to the full data set corresponding to $33.0$ kt-yr of
data of the Super-Kamiokande experiment as well as to all other
experiments in order to compare the active-active and active-sterile 
neutrinos oscillation channels to the atmospheric neutrino anomaly.

\end{abstract}

\maketitle

Atmospheric showers are initiated when primary cosmic rays hit the
Earth's atmosphere. Secondary mesons produced in this collision,
mostly pions and kaons, decay and give rise to electron and muon
neutrino and anti-neutrinos fluxes \cite{review}. In the past 
Fr\'ejus and NUSEX~\cite{atmexp-} reported a
R-value ($R=(\mu/e)_{\rm data}/(\mu/e)_{\rm MC}$) 
consistent with one, therefore other detectors like 
Kamiokande, IMB and Soudan-2~\cite{atmexp+} have measured $R$
significantly smaller than unity. 
Recent Super-Kamiokande high statistics
observations~\cite{SuperKamiokande} indicate that the deficit in the
ratio $R$ is due to the number of neutrinos arriving to the
detector at large zenith angles.

The main aim of this talk is to compare the active-active and active-sterile 
neutrinos oscillation channels to the atmospheric neutrino anomaly using 
the the new sample of $33.0$ kt-yr of Super-Kamiokande
data. This analysis uses the latest improved calculations of the 
atmospheric neutrino fluxes as a function of zenith angle, 
taking into account a variable neutrino production point~\cite{flux}.  

The expected neutrino event number both in the
absence and the presence of oscillations can be written as:

\begin{eqnarray}
& N_\alpha= & n_t T\sum_\beta 
\int \kappa_\alpha
\frac{d^2\Phi_\alpha}{dE_\nu d(\cos\theta_\nu)}  P_{\alpha\beta} 
\frac{d\sigma}{dE_\beta} \times  \nonumber \\
& & \varepsilon(E_\beta) dE_\nu dE_\beta d(\cos\theta_\nu)dh\; ;
\label{eventsnumber}
\end{eqnarray}

\noindent and $P_{\alpha\beta}$ is the transition probability for
$\nu_\alpha \to \nu_\beta$, 
$ P_{\alpha\beta} \equiv P(E_{\nu}, \cos\theta_\nu, h) $, 
where $\alpha,\beta=\mu ,e$.  
In the case of no oscillations, 
$P_{\alpha\alpha}=1$ for all $\alpha$.

Here $n_t$ is the number of targets, $T$ is the experiment's running
time, $E_\nu$ is the neutrino energy and $\Phi_\alpha$ is the flux of
atmospheric neutrinos of type $\alpha$; $E_\beta$ is the final
charged lepton energy and $\varepsilon(E_\beta)$ is the detection
efficiency for such charged lepton; $\sigma$ is the neutrino-nucleon
interaction cross section, and $\theta_\nu$ is
the zenith angle;
$h$ is the slant distance and 
$\kappa_\alpha$
is slant distance distribution~\cite{flux}.

We assume a two-flavor oscillation scenario, i. e. 
the $\nu_\mu$ oscillates into another flavour either $\nu_\mu \to
\nu_e$ , $\nu_\mu \to \nu_s$ or $\nu_\mu \to
\nu_\tau$~\cite{ourwork}. The evolution equations of 
the $\nu_\mu -\nu_X$ system
(where $X=e,\tau $ or $s$ sterile)  in the matter is
\begin{eqnarray}
i{\mbox{d} \over \mbox{d}t}\left(\matrix{
\nu_\mu \cr\ \nu_X\cr }\right) & = & 
 \left(\matrix{
   0
& {H}_{\mu X} \cr
 {H}_{\mu X} 
& {H}_X \cr}
\right)
\left(\matrix{
\nu_\mu \cr\ \nu_X \cr}\right) \,\,, \\
H_X   & \! = & \! 
V_X-V_\mu - \frac{\Delta m^2}{2E_\nu} \cos2 \theta_{\mu X}, \nonumber \\
H_{\mu X}& \!= &  - \frac{\Delta m^2}{4E_\nu} \sin2 \theta_{\mu X} \nonumber
\label{evolution1}
\end{eqnarray}
%
Here $\Delta m^2=m_2^2-m_1^2$. If $\Delta m^2>0 \: (\Delta m^2<0)$ 
the neutrino with largest
muon-like component is heavier (lighter) than the one with largest
X-like component. The functions $V_X$ are the usual matter potentials. 
In order to obtain the oscillation probabilities
$P_{\alpha\beta}$ we have made a numerical integration of the
evolution equation.  Notice
that for the $\nu_\mu\to\nu_\tau$ case there is no matter effect while
for the $\nu_\mu\to\nu_s$ case we have two possibilities depending on
the sign of $\Delta m^2$.  

\begin{figure}[hbt]
\vskip -1.0cm
\centerline{\protect\hbox{\epsfig{file=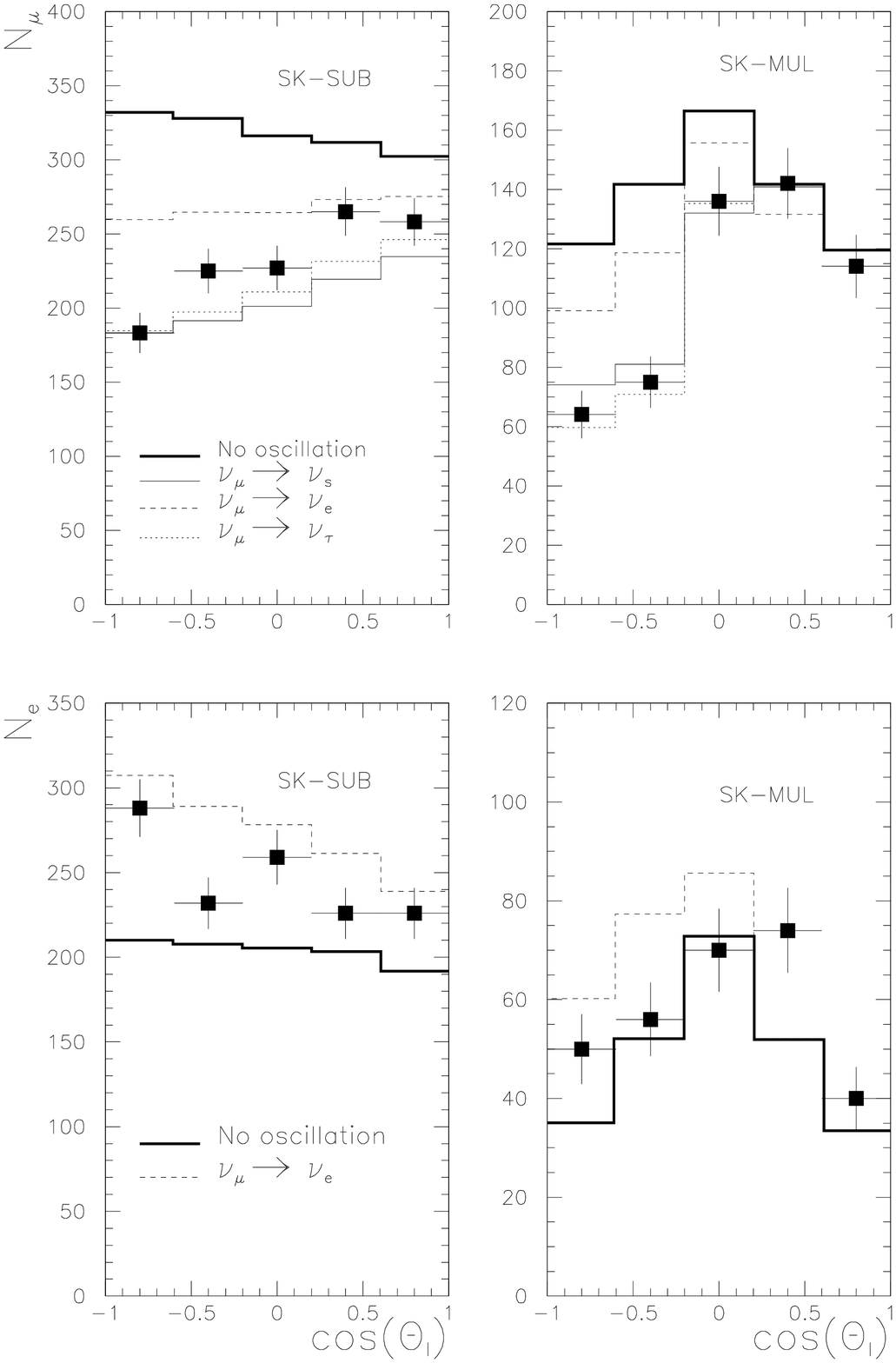,width=0.55\textwidth,height=0.51\textheight}}}
\vskip -1.18cm
\caption{
Angular distribution for Super-Kamiokande electron-like and 
muon-like sub-GeV and multi-GeV events, together with the prediction in the
absence of oscillation (thick solid line) as well as the prediction
for the best fit point for  $\nu_\mu \to
\nu_s$ (thin solid line), $\nu_\mu \to \nu_e$ (dashed line) and
$\nu_\mu \to \nu_\tau$ (dotted line) channels.
}
\label{ang_mu}  
\end{figure}

The steps required in order to generate the allowed regions of
oscillation parameters were described in Ref.~\cite{ourwork}. 
The $\chi^2$ is defined as 
\begin{equation}
\chi^2 \equiv \sum_{I,J}
X_I \cdot (\sigma_{data}^2 + \sigma_{theory}^2)_{IJ}^{-1}\cdot X_J,
\label{chi2}
\end{equation}
where $I = (A, \alpha)$ and $J = (B,
\beta)$ where, $A,B$ stands for Fr\'ejus, Kamiokande, IMB,... and
$\alpha, \beta = e,\mu$.  In Eq.~(\ref{chi2}) $N_I^{theory}$ is the
predicted number of events calculated from Eq.~(\ref{eventsnumber})
whereas $N_I^{data}$ is the number of observed events. The vector
$X_I$ is defined as $X_I\equiv N_I^{data}-N_I^{theory}$.
In Eq.~(\ref{chi2}) $\sigma_{data}^2$ and $\sigma_{theory}^2$ are the
error matrices containing the experimental and theoretical errors
respectively. The error matrices can be written 
$(\sigma_{IJ})^2 \equiv \sigma_\alpha(A)\, \rho_{\alpha \beta} (A,B)\,
\sigma_\beta(B)$ where $\rho_{\alpha \beta} (A,B)$ stands for the 
correlation between the $\alpha$-like events in the $A$-type experiment 
and $\beta$-like events in $B$-type experiment, whereas $\sigma_\alpha(A)$ 
and $\sigma_\beta(B)$ are the errors for the number of $\alpha$ and
$\beta$-like events in $A$ and $B$ experiments, respectively. 
The computation of correlations and errors are described 
in the Refs.~\cite{ourwork,fogli2}. 

The following step is the minimization of the 
$\chi^2$ function in Eq.~(\ref{chi2}) and the determination the
allowed region in the $\sin^22\theta-\Delta m^2$ plane, for a given
confidence level, defined as $\chi^2 \equiv \chi_{min}^2  + 4.61\ (9.21)\   \  
\  \mbox{for}\  \ 90\  (99) \% \ \  \mbox{C.L.}$

\begin{figure}[h]
\vskip -0.8cm
\centerline{\protect\hbox{\epsfig{file=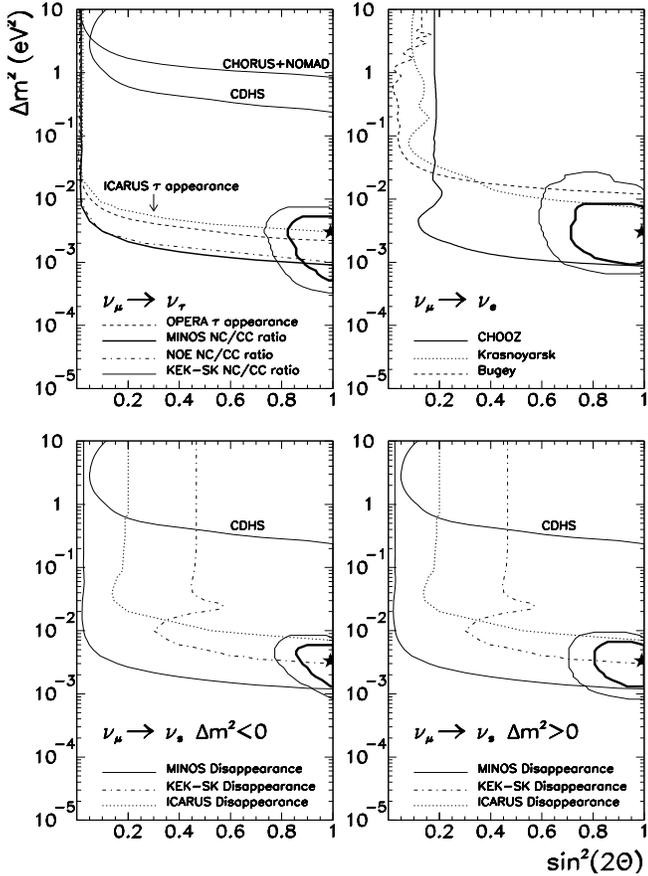,width=0.6\textwidth,height=0.62\textheight}}}
\vskip -1.5cm
\caption{
Allowed oscillation parameters for all experiments combined at 90
(thick solid line) and 99 \% CL (thin solid line) for each oscillation
channel as labeled in the figure.  
The best fit point is marked with a star.}
\label{mutausk4} 
\end{figure}

\newpage 

The results of our $\chi^2$ fit of the Super-Kamiokande sub-GeV and
multi-GeV atmospheric neutrino data can be 
appreciated in Ref.~\cite{ourwork}. It is possible to see the discrimination 
power of atmospheric neutrino data looking for the 
predicted zenith angle distributions for the various oscillation channels. 
As an example we take the case
of the Super-Kamiokande experiment and compare separately the sub-GeV
and multi-GeV data with what is predicted in the case of
no-oscillation and in all oscillation channels
for the corresponding best fit points obtained for the {\sl combined}
sub and multi-GeV data analysis. This is shown in Fig.~\ref{ang_mu}.

I now turn to the comparison of the information obtained from the
analysis of the atmospheric neutrino data with the
results from reactor and accelerator experiments as well as the
sensitivities of future experiments. For this purpose I present the
results obtained by combining all the experimental atmospheric
neutrino data from various experiments~\cite{atmexp-,atmexp+,SuperKamiokande}.
In Fig.~\ref{mutausk4} we show the combined information obtained from
the analysis of all atmospheric neutrino data 
and compare it with the constraints from
reactor experiments such as Krasnoyarsk, Bugey, and CHOOZ~\cite{reactors}, 
and the accelerator experiments such as CDHSW, CHORUS, and 
NOMAD~\cite{accelerators}. We also include in the same figure the sensitivities
that should be attained at the future long-baseline experiments now
under discussion~\cite{chiaki}.

To conclude we find that the regions of oscillation
parameters obtained from the analysis of the atmospheric neutrino data 
cannot be fully tested by the LBL experiments, when the 
Super-Kamiokande data are included in the fit for the $\nu_\mu \to
\nu_\tau$ channel as can be seen clearly from the upper-left panel
of Fig.~\ref{mutausk4}. One important point is that from the upper-right 
panel of Fig.~\ref{mutausk4} one sees that the CHOOZ reactor
data already exclude completely the allowed region for the $\nu_{\mu} \to
\nu_e$ channel when  all experiments are  combined at 90\% CL. For the 
sterile neutrino case most of the LBL experiments can not completely
probe the region of oscillation parameters indicated by this analysis, 
even with $\Delta m^2<0$ ( or  with $\Delta m^2>0$) respectively 
the lower-left panel (lower-right panel) in Fig.~\ref{mutausk4}.


\end{document}